\documentclass[10pt,a4]{article}

\def\be{\begin{equation}}
\def\ee{\end{equation}}
\def\bc{\begin{center}}
\def\ec{\end{center}}

\usepackage{}

\begin{document}
\begin{titlepage}

\title{One-parameter family of additive energies and momenta in 1+1 dimensional STR}

\author{Marek Pawlowski\\Soltan Institute for Nuclear Studies\\Hoza 69, 00-681 Warsaw, Poland\\e-mail: pawlowsk@fuw.edu.pl}

\date{}

\maketitle

\begin{abstract}

The velocity dependence of energy and momentum is studied. It is
shown that in the case of STR in the space-time of only one spatial
dimension the standard energy and momentum definition can be
naturally modified without lost of local Lorenz invariance,
conservation rules and additivity for multiparticle system. One
parameter family of energies and momenta is constructed and it is
shown that within natural conditions there is no further freedom.
Choosing proper family parameter one can obtain energy and momentum
increasing with velocity faster or slower in comparison with the
standard case, but almost coinciding with them in the wide velocity
region.

\end{abstract}

\end{titlepage}

\section{}
It is generally believed that the structure of space-time at Planck
scale may differ from our everyday experience. Quantum phenomena,
including quantum gravity, are expected to modify local space-time
structure. Such modification can break local Lorenz invariance, can
modify it or leave it untouched. A class of models exploring the
last possibility is known as Double Special Relativity (DSR)
\cite{dsr}. Their authors, being motivated by theoretical arguments
and some experimental question marks \cite{cosmic}, introduce
additional scale (connected with length or energy) and assume that
the local Lorenz invariance is preserved. They pay the price of
loose of standard definition of momentum and energy. In fact the
price is even higher: momentum is no longer an additive quantity
(momentum of a system is not a sum of momenta of its components) in
DSR models; the same concerns energy \cite{kosinski}. The reason is
clear. Simple textbook arguments \cite{Feynman} lead to the
conclusion that the standard STR definition of momentum:
\be\label{p-einstein}
{\bf p}(m, {\bf v}) = \frac{m{\bf
v}}{\sqrt{1-v^2/c^2}} \ee is the unique Lorenz covariant function of
$\bf v$ being additive and satisfying reasonable set of conditions.
Similarly the energy function is also unique: \be\label{e-einstein}
E(m, {\bf v}) = \frac{mc^2}{\sqrt{1-v^2/c^2}}. \ee

We show below that the standard arguments are not sufficient to
select (\ref{p-einstein},\ref{e-einstein}) in 1+1-dimensional
Minkowski space. The admissible form of momentum and energy is more
general and depends on a new arbitrary dimensionless parameter.

\section{}

It is well known that the one dimensional relativistic velocity
group $(V^1,\oplus)$ is commutative and associative (in contrast to
the case of dimension 3). In fact the group is isomorphic with
ordinary $(R,+)$ group. The group isomorphism can be written in
example in the following form:

\be\label{isomorphism} V^1 \ni v \rightarrow
y(v)=\log{\frac{1+v}{1-v}} \in R \ee

The one dimensional velocity addition is given by
\be\label{v-composition} v \oplus u = \frac{v+u}{1+vu}, \ee
(we have
assumed $c=1$ for simplicity)

Its counterpart takes the simple "linear" form
\be\label{y-composition} y(v \oplus u) = y(v) + y(u). \ee

Observe the following properties: \be\label{zero} y(0) = 0, \ee
\be\label{minus} y(-v) = -y(v). \ee

These observations will be very useful for our subsequent
considerations.

\section{}

Our aim is to define additive relativistic momentum.
\medskip

Assume that there exists the momentum which is a real function
$p(m,v)$ that can be ascribed to each separate body (of low energy
defined mass $m$) and that fulfils the following conditions:
\begin{enumerate}
 \item $p(m,-v)=-p(m,v)$ (antisymmetry)
 \item For a two body elastically scattering system we define
 the center of mass reference frame $U_{CM}$ (it is the frame
 in which the asymptotic relations hold: $v^{in}_I =v^{CM}_I =
 -v^{out}_I$, $I=1,2$). We'll demand that the sum of momenta vanishes
 in $U_{CM}$:
\be\label{CM} p(m_1,v^{CM}_1)+p(m_2,v^{CM}_2)=0. \ee
 \item There holds the relativistic invariant momentum conservation
 rule:
 \begin{eqnarray}\nonumber p(m_1,v^{CM}_1 \oplus u) +
 p(m_2,v^{CM}_2 \oplus u)=\\ \label{momentumconservation}
 p(m_1,-v^{CM}_1 \oplus u)+p(m_2,-v^{CM}_2 \oplus u)
 \end{eqnarray} for arbitrary $u$ and arbitrary set of $m_I, v_I^{CM}$ satisfying (\ref{CM}).
 \item $p(m,v)\rightarrow mv$ for $v\ll 1$ (correspondence
 principle).
 \end{enumerate}

We will show that the most general function satisfying conditions
1.-4. is \be\label{momentum}
p(m,v)=\frac{m}{a}\sinh\left(\frac{a}{2}
\log{\frac{1+v}{1-v}}\right) \ee where $a$ is an arbitrary real or
pure imaginary parameter.
\smallskip

Observe that by means of the isomorphism (\ref{isomorphism}) the
momentum (\ref{momentum}) can be concisely rewritten:
\be\label{y-momentum} p(m,y)=\frac{m}{a}\sinh(\frac{ay}{2}). \ee

\medskip

The proof of (\ref{momentum}) goes as follows:
\medskip

Using antisymmetry of $p$ and $y$ one can rewrite
(\ref{momentumconservation})
\begin{eqnarray}\nonumber p(m_1,y_1 + w) - p(m_1,y_1 - w) +
\\ \label{proof1} p(m_2,y_2 + w) + p(m_2,y_2 - w) =0.
 \end{eqnarray} where $w=y(u)$.
Differentiating (\ref{proof1}) $2n$ times with respect to the second
argument and putting $w=0$ we get the set of relations
\be\label{proof2} p^{(2n)}(m_1,y_1)+p^{(2n)}(m_2,y_2)=0. \ee Let us
treat the first of them (for $n=0$) as an involved relation between
$y_1$ and $y_2$. Now, differentiating it once and twice with respect
to $y_1$ and combining the results with (\ref{proof2}) for $n=1$ (in
order to eliminate the second body momentum) we get
\be\label{proof3}
\frac{d}{dy_1}\left(\log\frac{dp(m_1,y_1)}{dy_1}\right)=\frac{\frac{d^2
y_2}{dy_1^2}}{\frac{dy_2}{dy_1}\left(
1-\left(\frac{dy_2}{dy_1}\right)^2\right)}. \ee The similar equation
but with the third derivative under the logarithm, can be obtained
on the similar way starting from $n=1$. As the RHS of the both
formulas are identical, we get the differential equation
\be\label{proof4} \frac{d}{dy}\left(\log\frac{dp(m,y)}{dy}\right)=
\frac{d}{dy}\left(\log\frac{d^3p(m,y)}{dy^3}\right) \ee where the
subscripts have been omitted.

Applying again the conditions 1. and 4. in order to fix some
integration constants we get the general solution of (\ref{proof4})
in the form given by (\ref{y-momentum}).
\medskip

The obtained momentum (\ref{momentum}) comes to the standard
expression (\ref{p-einstein}) for $a=1$. The plots of velocity
dependencies of (\ref{momentum}) for several values of $a$ are
compared on the Figure 1. If the value of $a$ is close to unity, the
dependencies are almost undistinguishable in the wide range of $v$.

The solution of (\ref{proof4}) admits also pure imaginary parameters
$a$. Some relevant plots are given on the Figure 2. The momenta with
imaginary $a$ are undistinguishable from the others for small
velocities. However their asymptotic behavior is bizarre.
\medskip

Despite the fact, that the momentum formula (\ref{momentum}) was
derived from considerations based on the elastic scattering, its
applicability is universal. Consider for example a two body decay.
Let $M$ be the decaying mass and $m_1, m_2$ be the masses of the
decay products. Then assuming momentum conservation principle  and
its relativistic invariance we come to the conclusion that
\be\label{decay} M=m_1 \cosh(ay_1)+m_2 \cosh(ay_2). \ee We'll make
use of this relation deriving generalized energy formula.

\section{}

The generalized energy can be derived from the set of assumptions
similar to 1.-4. Energy is expected to be a symmetric function of
velocity (i), to be additive and conserved in every inertial
reference frame (ii) and to have a proper low energy limit (iii). In
addition it is demanded that energy is conserved in two-body decays
(iv).

A similar consideration as in the case of momentum leads to the
generic formula carrying out all the conditions: \be\label{energy}
E(m,v)=m\left((1-\frac{1}{a^2})+\frac{1}{a^2}\cosh\left(\frac{a}{2}\log\frac{1+v}{1-v}\right)\right)
\ee or in an equivalent notation \be\label{y-energy}
E(m,y)=m\left((1-\frac{1}{a^2})+\frac{1}{a^2}\cosh\left(\frac{ay}{2}\right)\right).
\ee It follows from the proof of (\ref{energy}) that the constant
$a$ is the same as previously. Again, for $a=1$ the energy
(\ref{energy}) takes its standard STR form (\ref{e-einstein}).
\medskip

The momentum and energy of a body given by (\ref{momentum}) and
(\ref{energy}) are connected by energy/momentum dispersion relation
\be\label{dispersion}
a^2\left(E-m(1-\frac{1}{a^2})\right)^2-p^2=\frac{m^2}{a^2} \ee which
is a simple consequence of the hypergeometric unity relation.

\section{}
The new dimensionless constant $a$ that appears naturally in energy
and momentum definitions in 1+1 dimensional space-time, allows us to
modify slightly the relation between energy/momentum and velocity.
For $a<1$ both the energy and momentum are reduced in high energy
limit in comparison with the standard definition. For $a>1$ they
increase. This is just the phenomenon that makes attractive DSR
models. In example it helps to explain eventual ultra high energy
events in cosmic radiation.

The presented analysis was done in a space-time of reduced
dimension. The analysis was strictly dependent on this dimension and
it is obvious that there is no straight way to generalize it. If we
want to obtain similar result in the physical space-time we have to
attenuate some natural conditions that usually are placed on
momentum and energy. Presented analysis shows a new type of
dispersion relation - in a sense alternative to the relations
studied in the literature in the context of DSR.

\eject Figure Captions
\bigskip

Fig. 1  $p(1,v)$ plots for $a=0.00001, 0.5, 0.7, 1, 1.2, 1.5$

Fig. 2  $p(1,v)$ plots for imaginary parameter $a=0.5 i, 0.8 i, i,
1.2 i, 1.5 i$ compared with $a=1$ and $a=0.00001$. (Plot range is
not symmetric.)

\eject

\begin{picture}(40,55)
\includegraphics{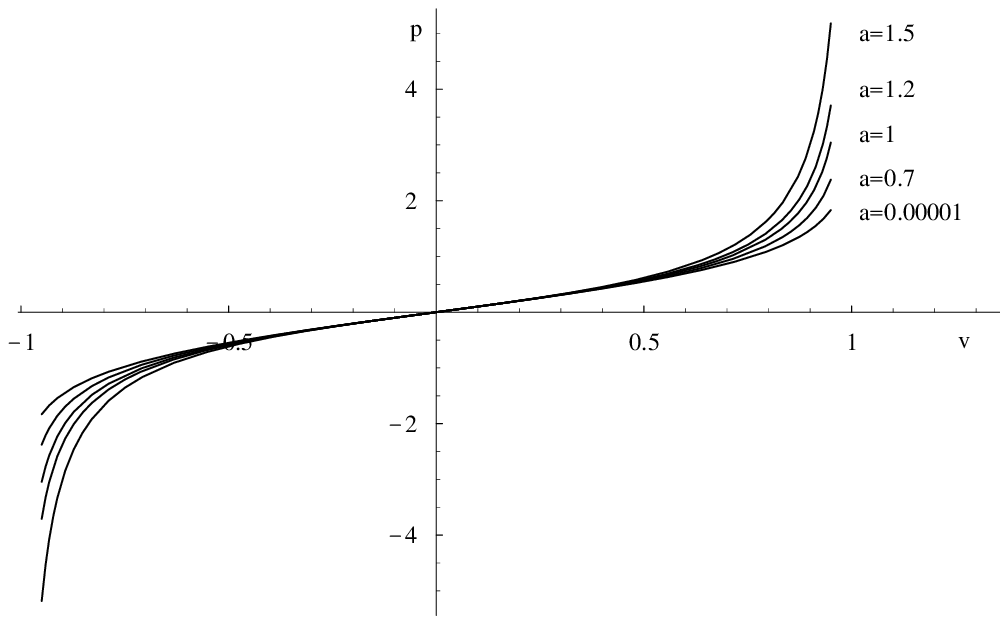} \put(0,-174){\centerline{{\bf Fig. 1}}}
\end{picture}

\eject
\begin{picture}(40,55)
\includegraphics{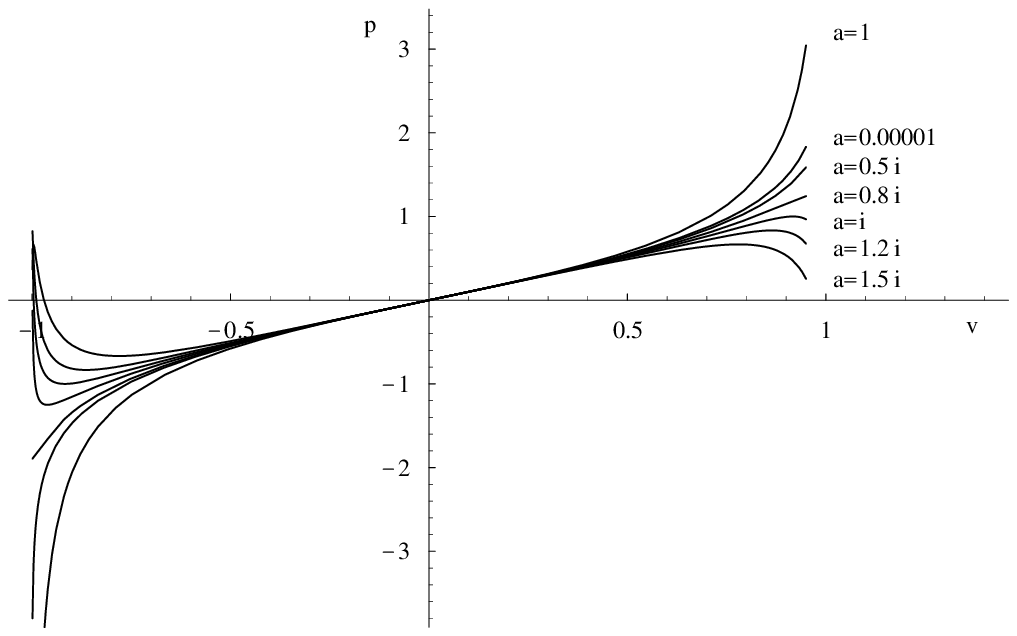} \put(0,-174){\centerline{{\bf Fig. 2}}}
\end{picture}

\end{document}